\documentclass{aa}
\usepackage{graphicx}
\usepackage{epsfig}
\usepackage{rotating}
\usepackage[T1]{fontenc}
\usepackage{amssymb}
\usepackage{ae,aecompl}
\usepackage{natbib}
\usepackage{natbib,twoopt}
\usepackage[breaklinks=true]{hyperref} %% to avoid \citet line fills

\bibpunct{(}{)}{;}{a}{}{,} %% natbib format for A&A and ApJ
\usepackage[varg]{txfonts}
\usepackage{graphicx}
\usepackage{color}           % For color text: \color command

\begin{document}

\title{Synthetic \textit{IRIS} spectra of the solar transition region:\\ Effect of high-energy tails}

\author{E. Dzif\v{c}\'{a}kov\'{a}\inst{1}, C. Vocks\inst{2} \and J. Dud\'{i}k\inst{1}}
\offprints{E.Dzif\v{c}\'{a}kov\'{a}; e-mail: elena@asu.cas.cz}
\institute{Astronomical Institute Academy of Sciences of the Czech Republic, 251 65 Ond\v{r}ejov, Czech Republic\\
             \email{elena@asu.cas.cz}
    \and
Leibniz-Institut für Astrophysik Potsdam, An der Sternwarte 16, 14482 Potsdam, Germany\\
               \email{cvocks@aip.de}}

\date{Received  / Accepted }

\abstract{}
{The solar transition region satisfies the conditions for presence of non-Maxwellian electron energy distributions with high-energy tails at energies corresponding to the ionization potentials of many ions emitting in the EUV and UV portions of the spectrum.}
{We calculate the synthetic \ion{Si}{IV}, \ion{O}{IV}, and \ion{S}{IV} spectra in the far ultra-violet (FUV) channel of the Interface Region Imaging Spectrograph (IRIS). Ionization, recombination, and excitation rates are obtained by integration of the cross-sections or their approximations over the model electron distributions considering particle propagation from the hotter corona.}
{The ionization rates are significantly affected by the presence of high-energy tails. This leads to the peaks of the relative abundance of individual ions to be broadened with pronounced low-temperature shoulders. As a result, the contribution functions of individual lines observable by \textit{IRIS} also exhibit low-temperature shoulders, or their peaks are shifted to temperatures an order of magnitude lower than for the Maxwellian distribution. The integrated emergent spectra can show enhancements of \ion{Si}{IV} compared to \ion{O}{IV} by more than a factor of two.}
{The high-energy particles can have significant impact on the emergent spectra and their presence needs to be considered even in situations without strong local acceleration.}


\titlerunning{Effect of high-energy tails on \textit{IRIS} spectra}
\authorrunning{Dzif\v{c}\'{a}kov\'{a}, Vocks, Dud\'{i}k}

\maketitle
\newpage

%__________________________________
\section{Introduction}
\label{Sect:1}

The solar transition region is an interface between the chromosphere and the multi-million Kelvin corona. Its temperatures of several times $10^4 - 10^5$\,K lead to emission of multiple ions such as \ion{C}{IV}, \ion{O}{IV}--\ion{O}{VI}, \ion{Si}{IV}, and others. The recent Interface Region Imaging Spectrograph \citep[\textit{IRIS},][]{DePontieu14} observes the \ion{Si}{IV}, \ion{O}{IV}, and \ion{S}{IV} emission in its far ultra-violet (FUV) channel at 1390–1407\,\AA. These observations typically show strong \ion{Si}{IV} doublet at 1393.8\,\AA~and 1402.8\,\AA~together with the neighboring \ion{O}{IV} multiplet, whose strongest \ion{O}{IV} 1401.2\,\AA~line is weaker by a factor of five or more than the \ion{Si}{IV} 1402.8\,\AA~one \citep[e.g.,][]{Doyle84,Judge95,Curdt01,Peter14Sci,Yan15,Polito16b,Doschek16}; this is despite the fact that the \ion{O}{IV} 1401.2\,\AA~line should be stronger than the \ion{Si}{IV} 1402.8\,\AA~one, if equilibrium Maxwellian distribution and photospheric abundances are assumed \citep{Dudik14a}.

By its very nature, the transition region is characterized by strong temperature gradients \citep[e.g.,][]{Dupree72,Fontenla90,Fontenla91,Fontenla93,Gudiksen05a,Gudiksen05b,Bradshaw13,Hansteen15}. Such conditions are favorable for occurrence of high-energy tails as a result of fast particles entering the transition region from the corona \citep[e.g.,][]{Roussel-Dupre80a,Shoub83,Ljepojevic88,Vocks16}. This behavior originates in the electron collision frequency being proportional to ${\cal E}^{-3/2}$, where ${\cal E}$ is the electron kinetic energy. \citet{Roussel-Dupre80a} showed that the high-energy tails in the electron and proton distributions can exist in the solar transition region. Furthermore, these high-energy tails occur at energies comparable to the ionization potential of ions emitting at transition-region temperatures in ionization equilibrium. This leads to changes in the ionization rates \citep{Roussel-Dupre80b,Shoub83} that can affect the line intensities arising in the transition region \citep[see also][]{Dudik14a}. Evidence for high-energy electrons in the transition region was recently obtained by \citet{Testa14}.

Following these work\textbf{s}, in a companion paper \citep[][hereafter Paper I]{Vocks16}, the electron distributions were obtained in the transition region below a closed coronal loop \citep[see also][]{Vocks08} 210\,Mm in length. Here, we use these numerical distributions to obtain the synthetic spectra emerging from the model transition region and compare these to the Maxwellian predictions. The model is described briefly in Sect. \ref{Sect:2}. The spectral synthesis procedure together with the results are subjects of Sect. \ref{Sect:3}. Section \ref{Sect:4} summarizes the results.

\begin{figure}
  \resizebox{\hsize}{!}{\includegraphics{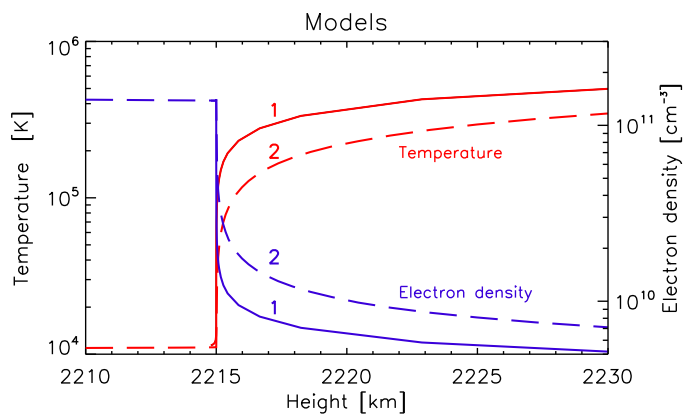}}
  \caption{Changes in the electron temperature ({\it red}) and density ({\it blue}) in two analyzed loop models. Numbers mark the model number.}
  \label{Fig:model}
\end{figure}

\begin{figure}
  \resizebox{\hsize}{!}{\includegraphics{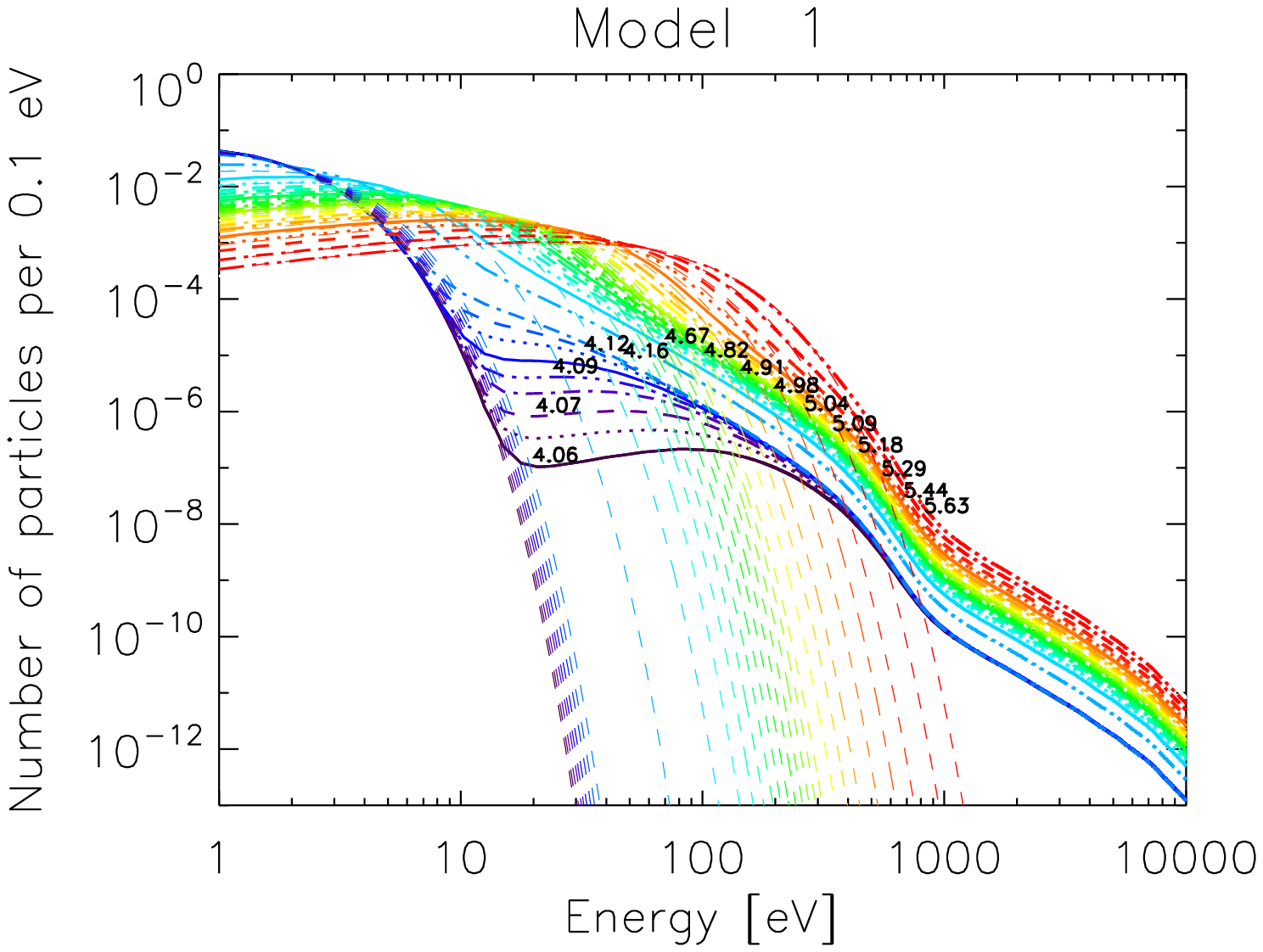}}
  \resizebox{\hsize}{!}{\includegraphics{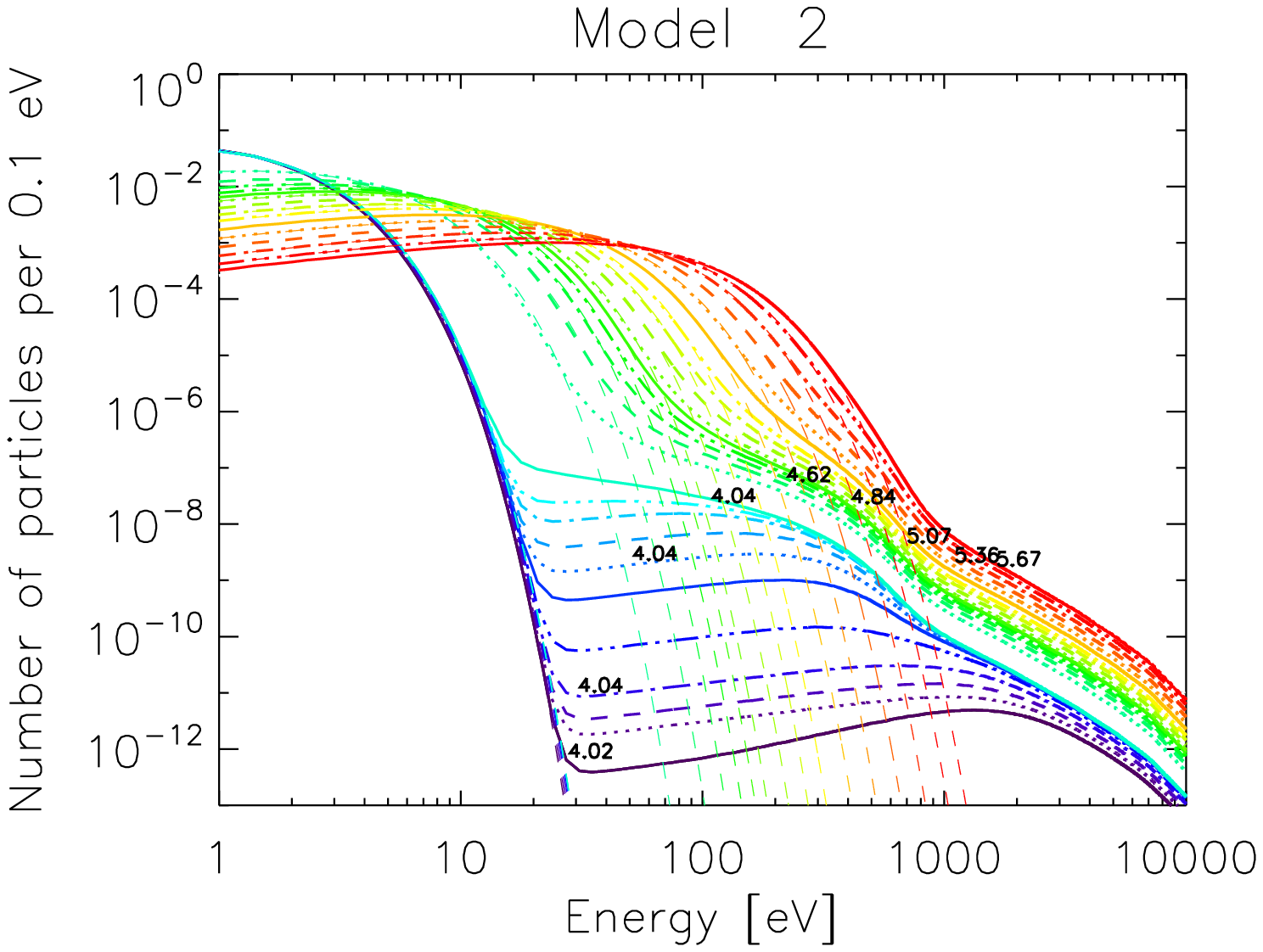}}
  \caption{Distribution functions at different positions in the transition-region along the loop in the Model 1 ({\it top}) and Model 2 ({\it bottom}). Different colors mark different distances along the loop model, and they are labeled by logarithm of temperature. Dashed lines correspond to the Maxwellian core of the distributions.}
  \label{Fig:distr}
\end{figure}

\begin{figure}
  \resizebox{\hsize}{!}{\includegraphics{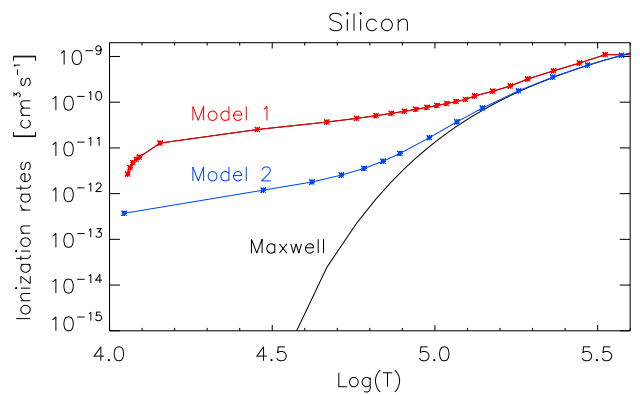}}
   \caption{Direct ionization rates as a function of temperature in the Model 1 ({\it red}), Model 2 ({\it blue}) and for the Maxwellian distribution ({\it black}). }
  \label{Fig:rates}
\end{figure}

\begin{figure}
  \resizebox{\hsize}{!}{\includegraphics{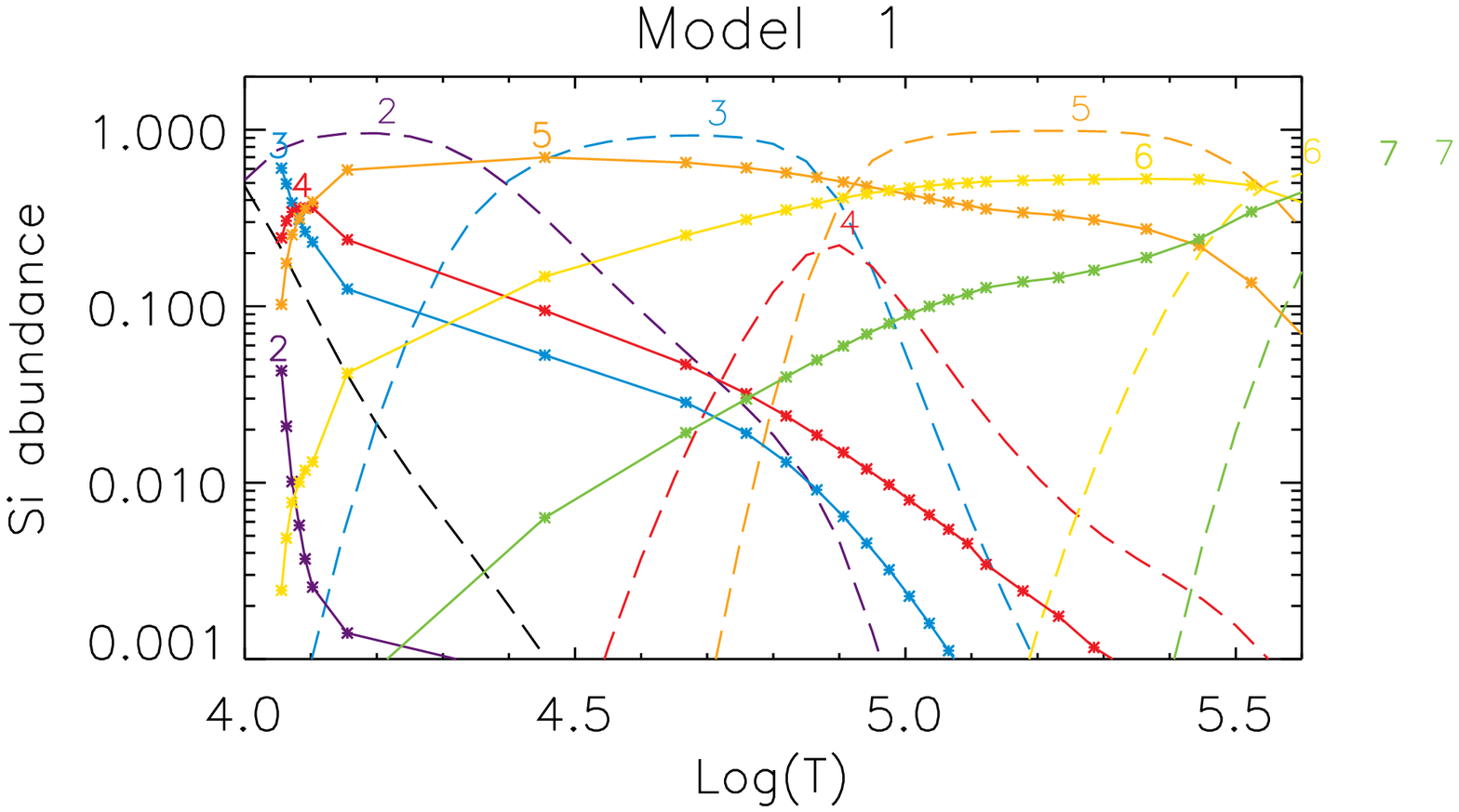}}
  \resizebox{\hsize}{!}{\includegraphics{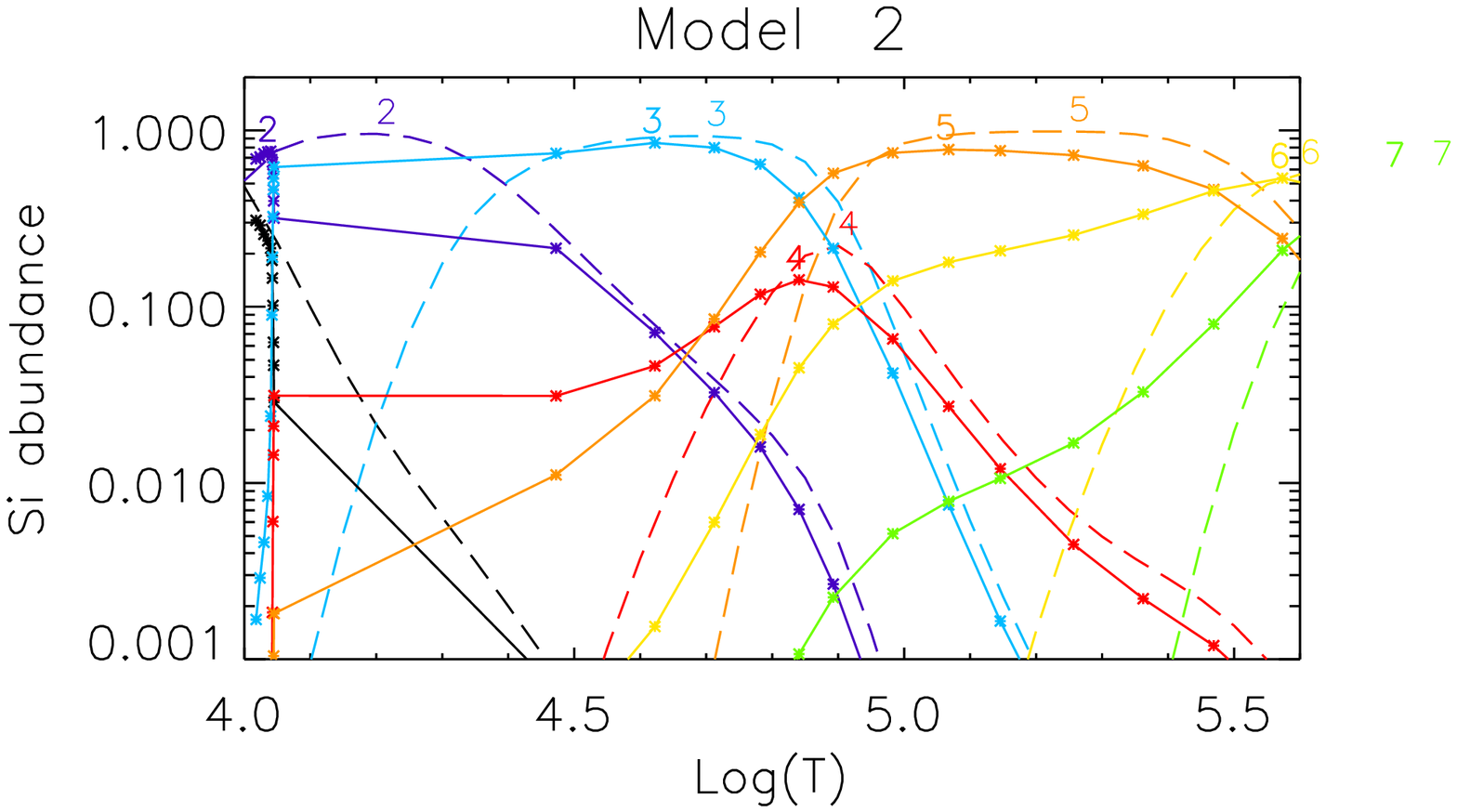}}
  \caption{Calculated Si ion abundances ({\it full lines}) for Model 1 ({\it top}) and Model 2 ({\it bottom}) in a comparison with the Maxwellian case ({\it dashed lines}). Different colors mark different ions and numbers correspond to the degree of ionization.}
  \label{Fig:ioneq_si}
\end{figure}

\begin{figure}
  \resizebox{\hsize}{!}{\includegraphics{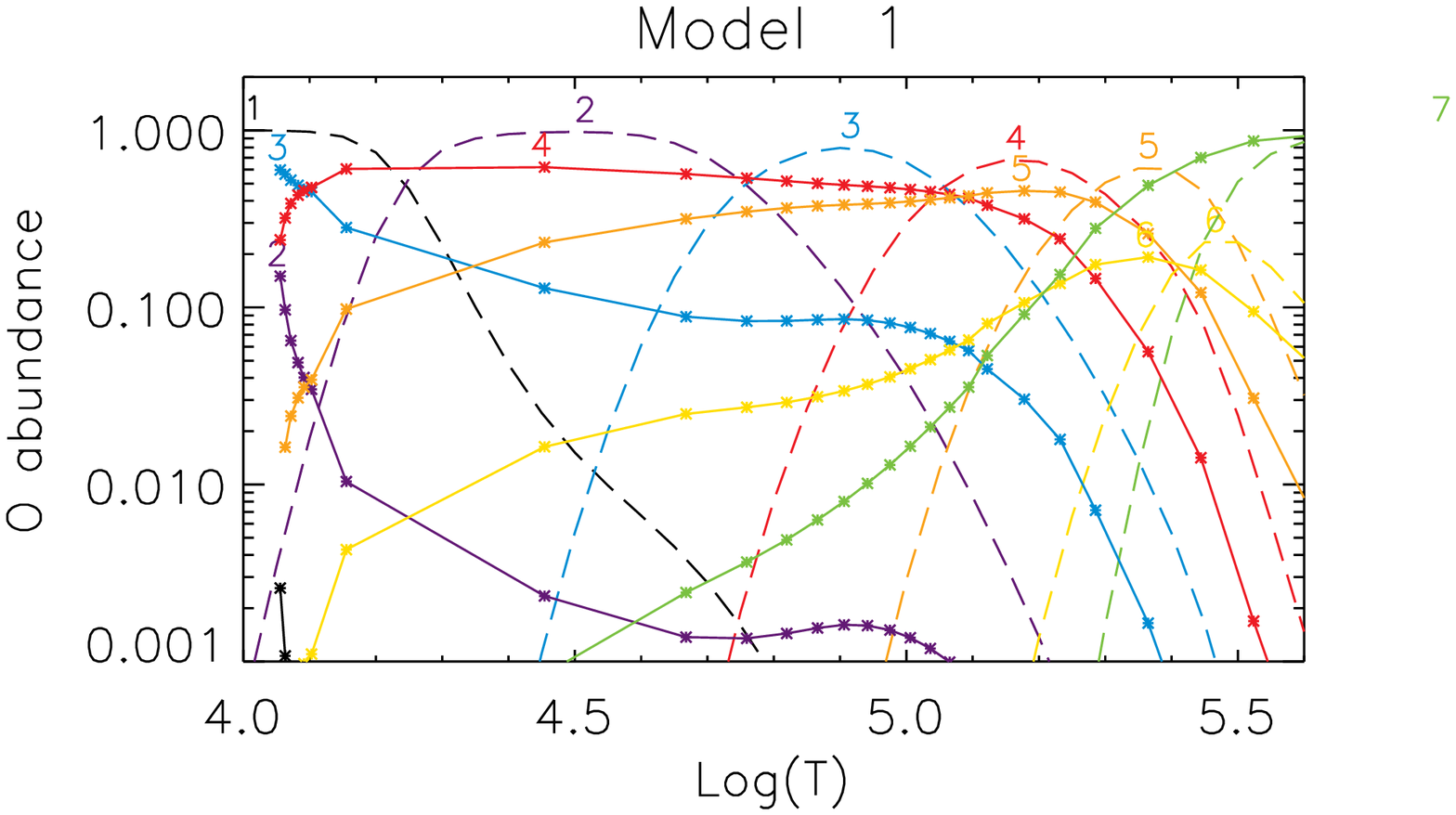}}
  \resizebox{\hsize}{!}{\includegraphics{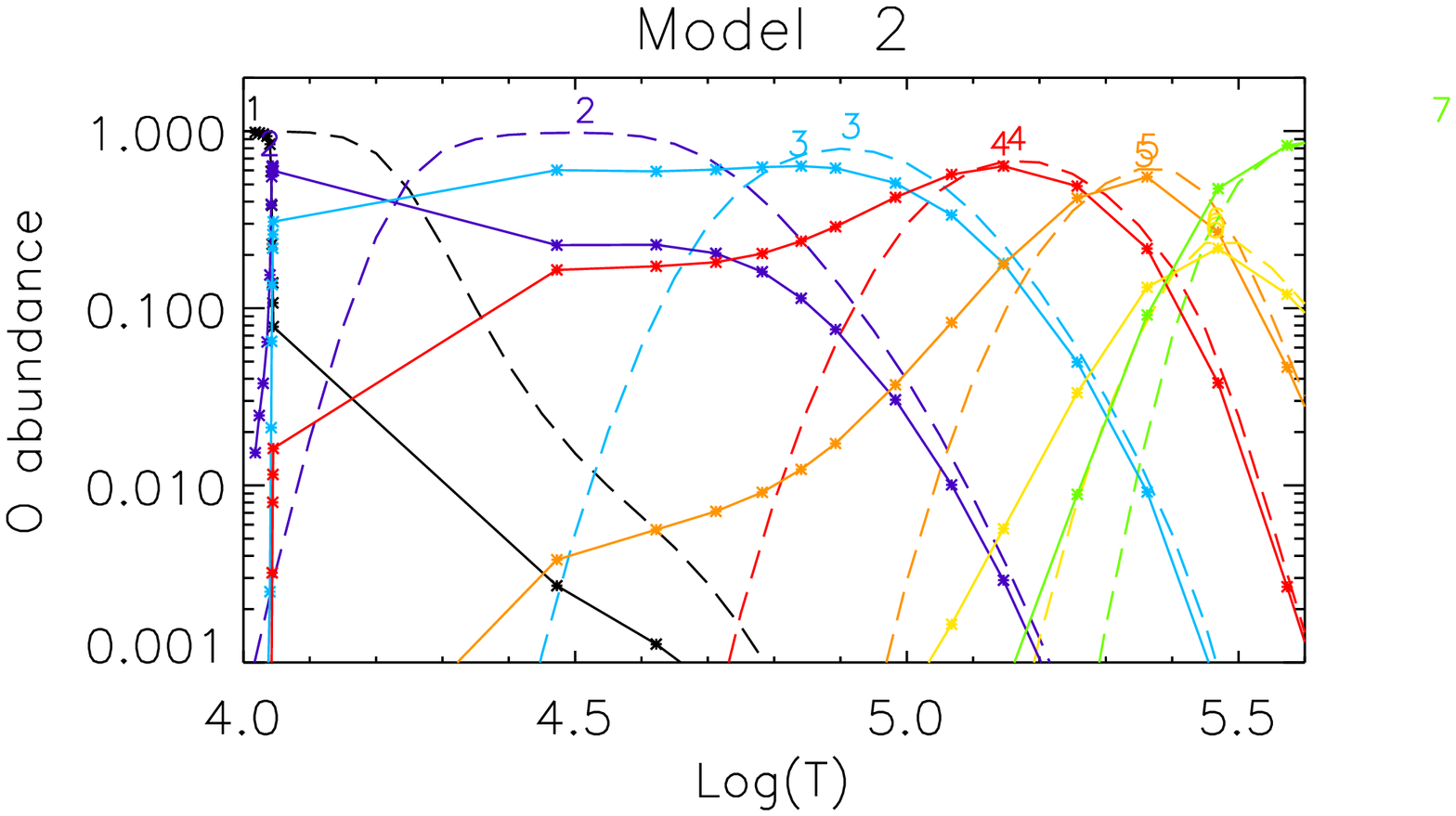}}
  \caption{Calculated O ion abundances ({\it full lines}) for Model 1 ({\it top}) and Model 2 ({\it bottom}) in a comparison with the Maxwellian case ({\it dashed lines}). Different colors mark different ions and numbers correspond to the degree of ionization.}
  \label{Fig:ioneq_o}
\end{figure}

\begin{figure}
  \resizebox{\hsize}{!}{\includegraphics{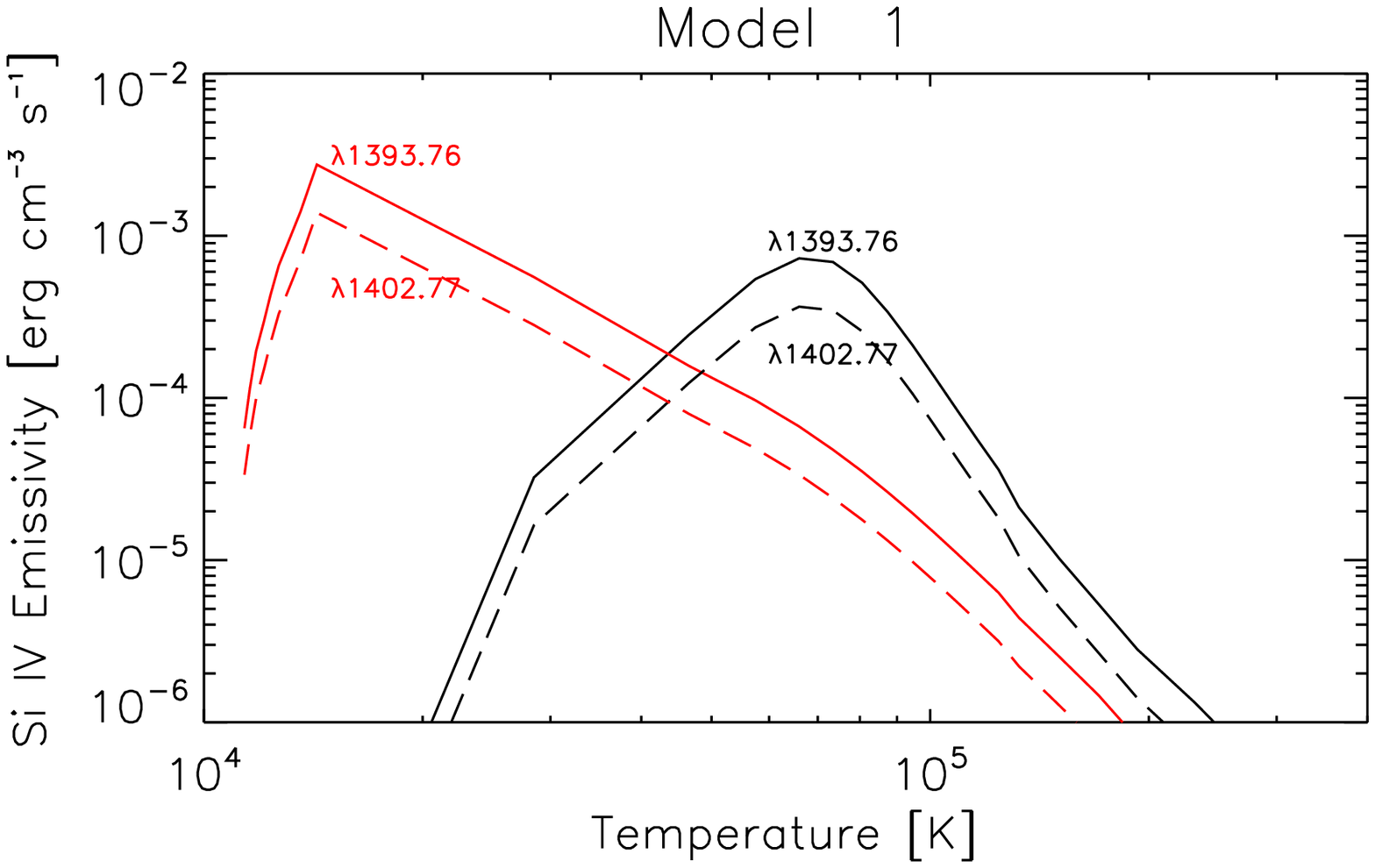}}
  \resizebox{\hsize}{!}{\includegraphics{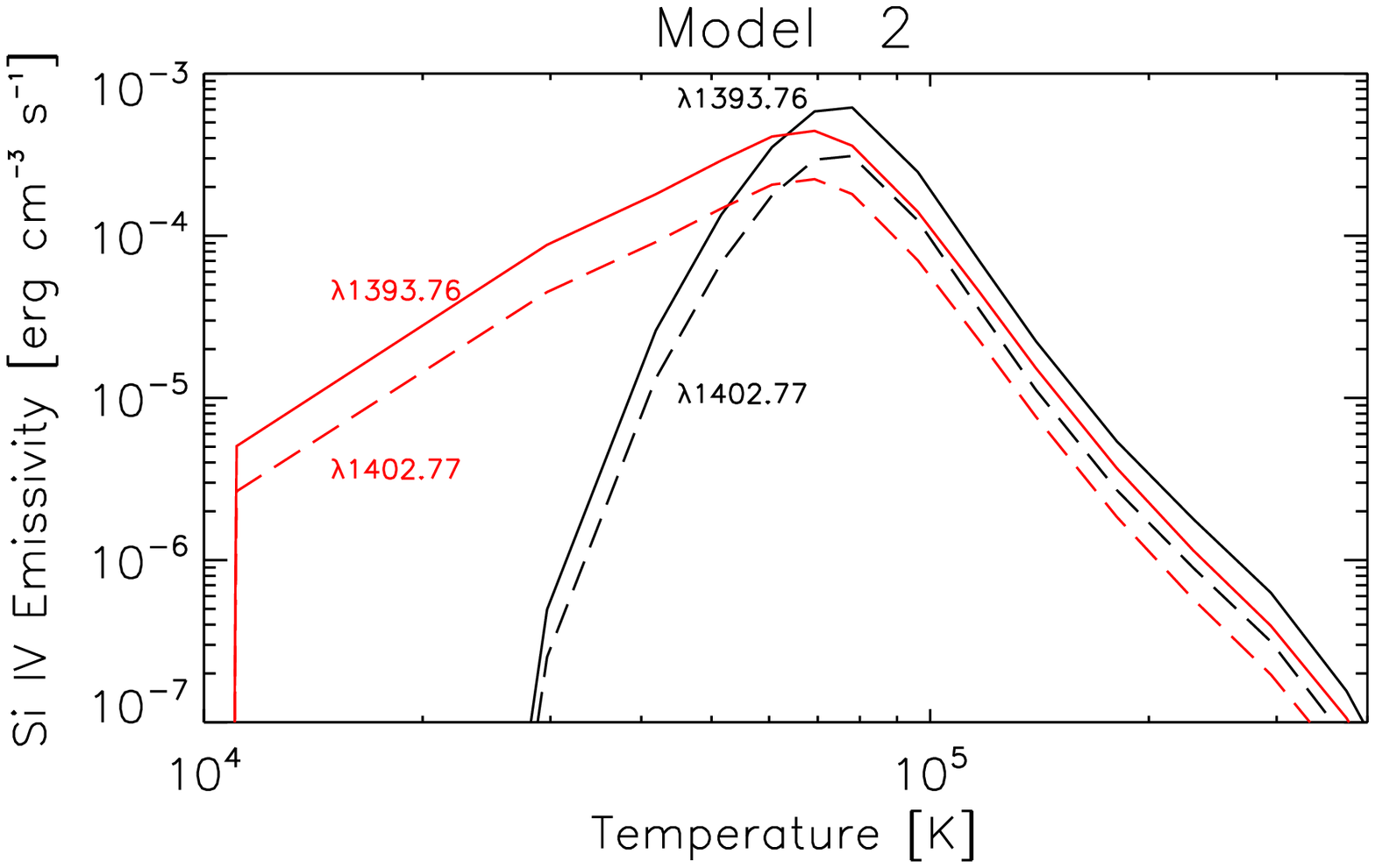}}
  \caption{Calculated emissivity of the strongest Si lines ({\it red lines}) for Model 1 ({\it top}) and Model 2 ({\it bottom}) in a comparison with the Maxwellian case ({\it black lines}). }
  \label{Fig:emiss_si}
\end{figure}

\begin{figure}
  \resizebox{\hsize}{!}{\includegraphics{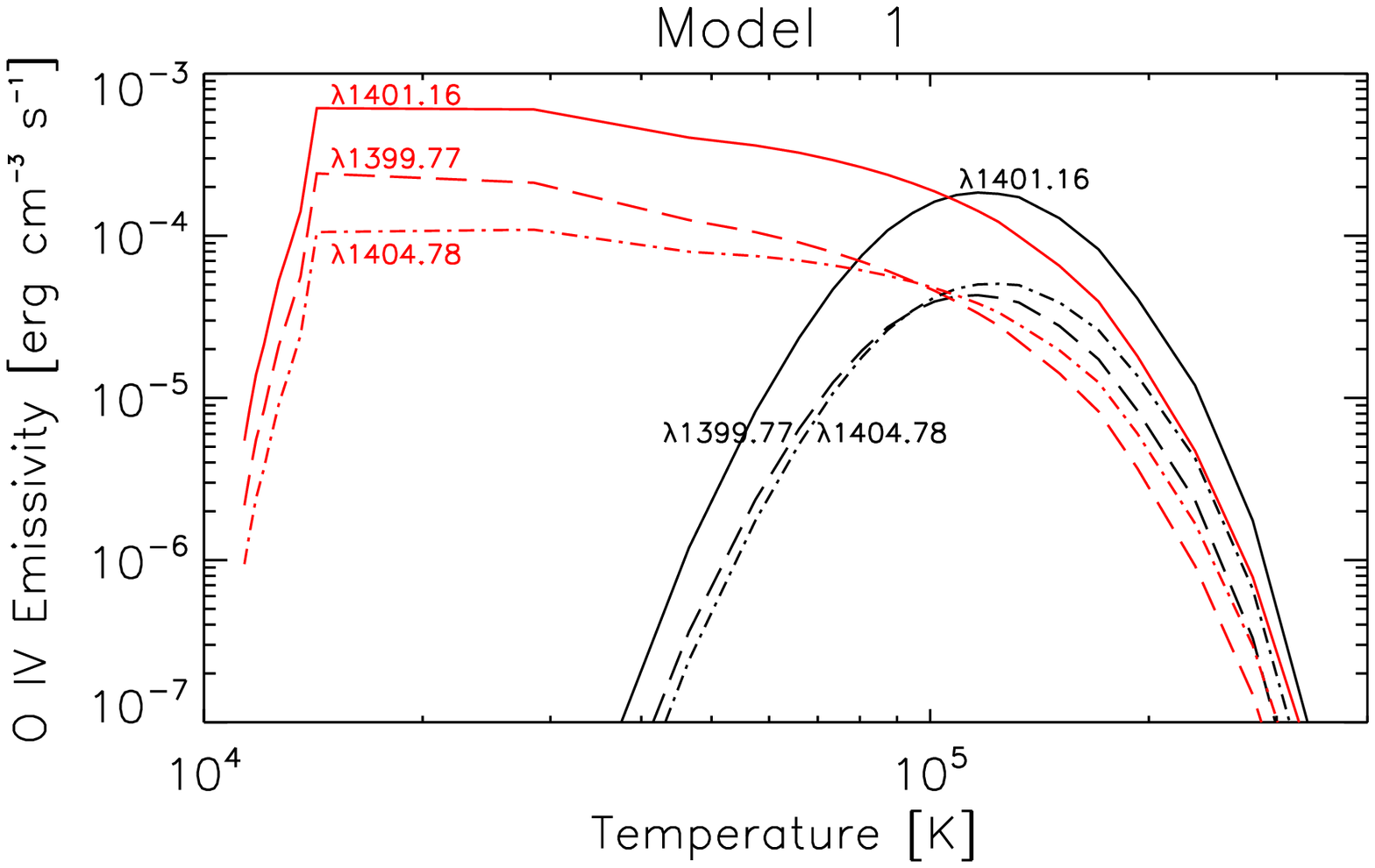}}
  \resizebox{\hsize}{!}{\includegraphics{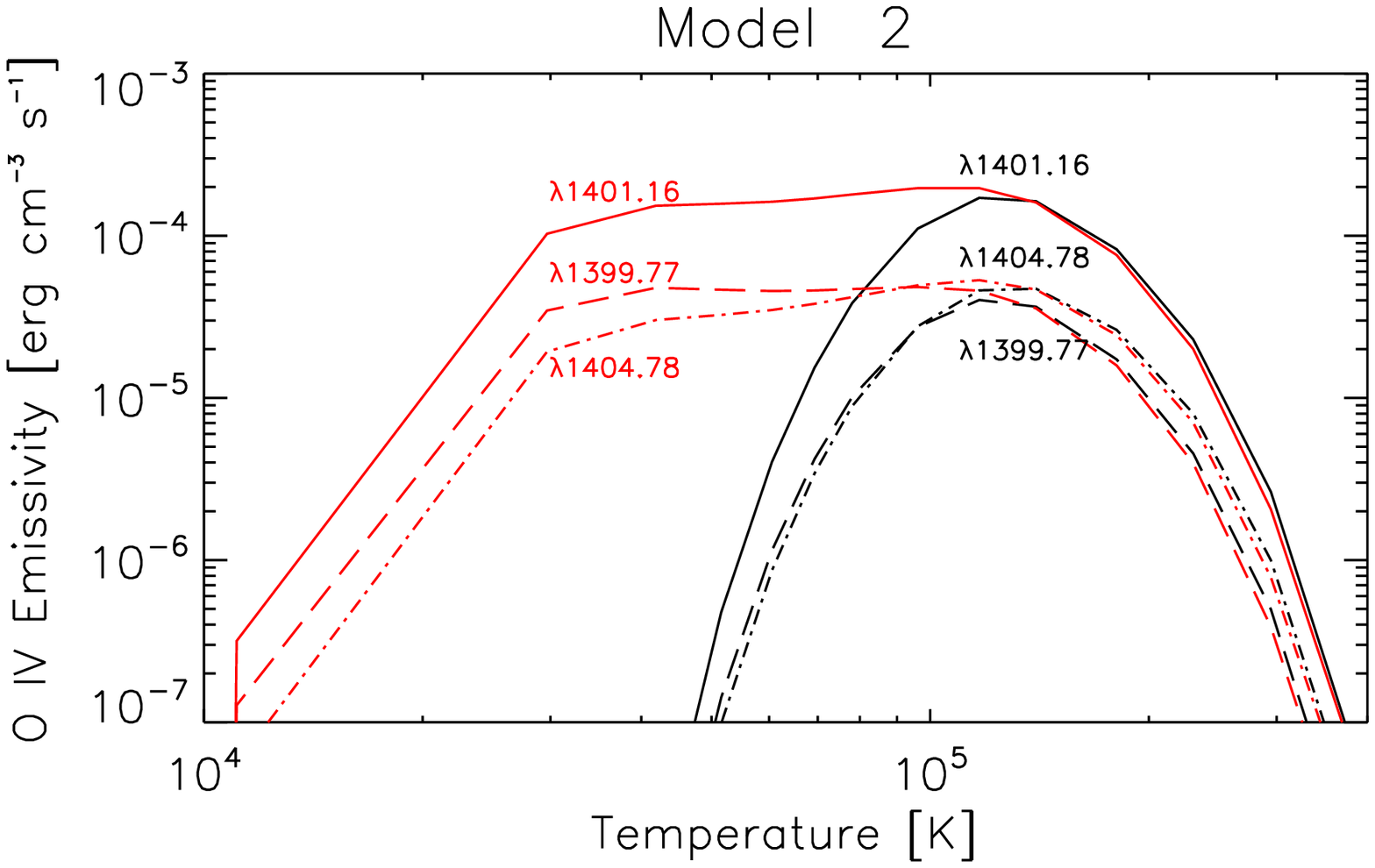}}
  \caption{Calculated emissivity of the strongest O lines ({\it red lines}) for Model 1 ({\it top}) and Model 2 ({\it bottom}) in a comparison with the Maxwellian case ({\it black lines}). }
  \label{Fig:emiss_o}
\end{figure}

\begin{figure}
  \resizebox{\hsize}{!}{\includegraphics{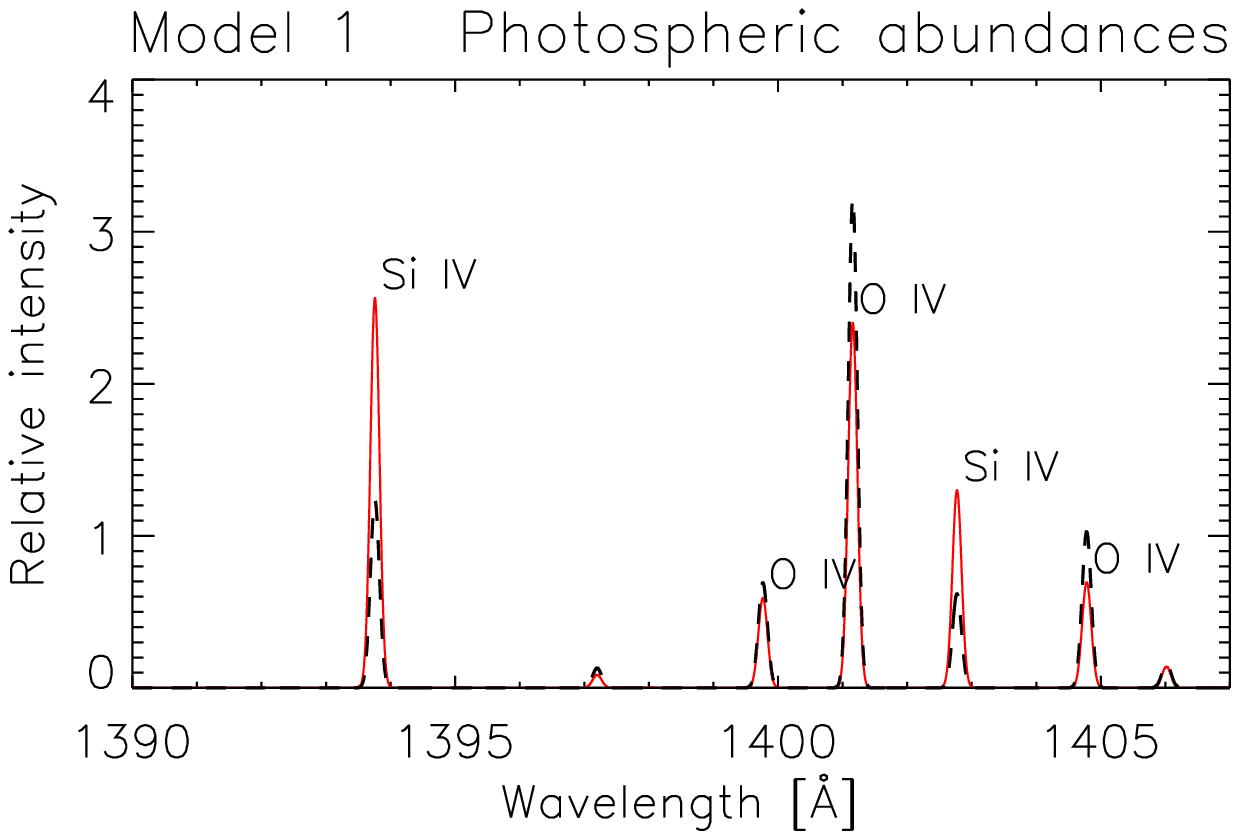}}
  \resizebox{\hsize}{!}{\includegraphics{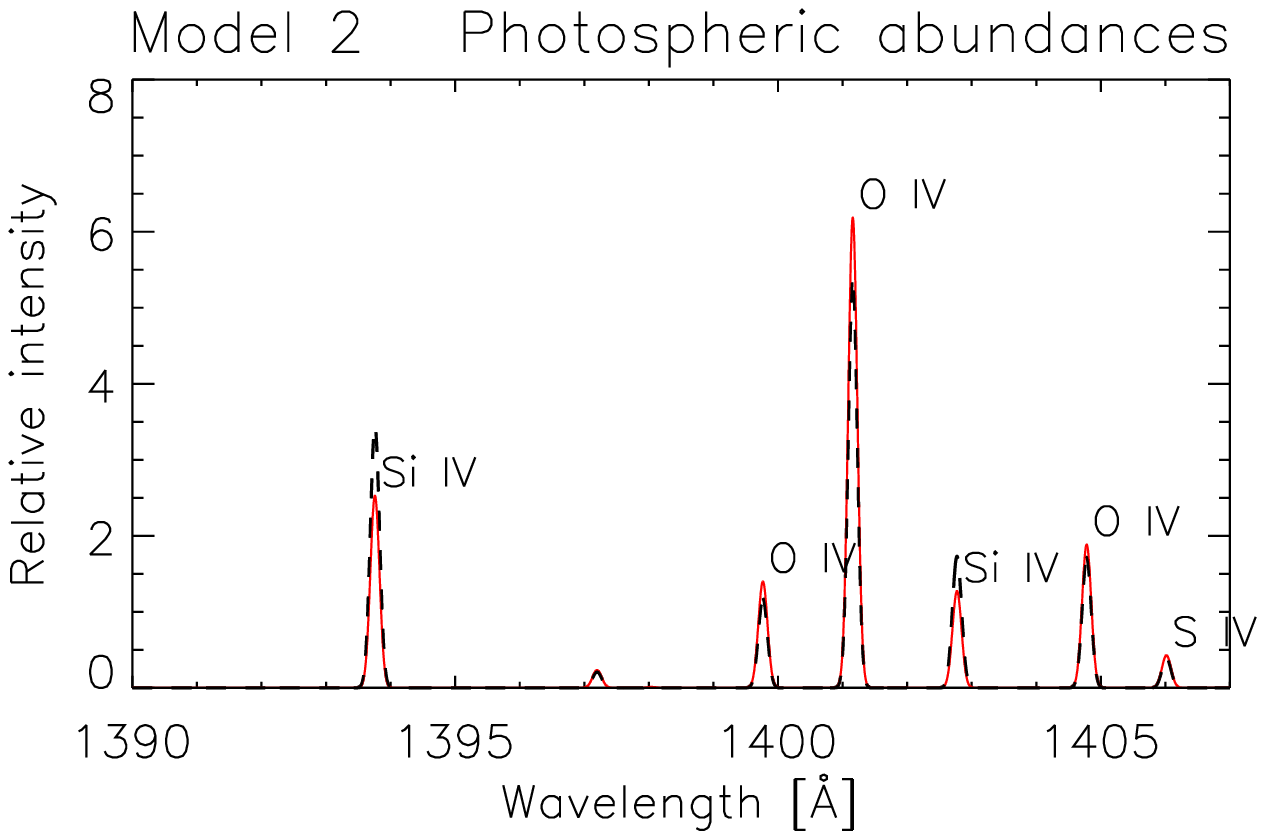}}
  \caption{Calculated synthetic spectrum observable in IRIS with the photospheric abundances ({\it full red lines}) for Model 1 ({\it top}) and Model 2 ({\it bottom}) in a comparison with the Maxwellian spectrum for the same temperature and density structure of the loop ({\it dashed black lines}). }
  \label{Fig:spectrum_pho}
\end{figure}

\begin{figure}
  \resizebox{\hsize}{!}{\includegraphics{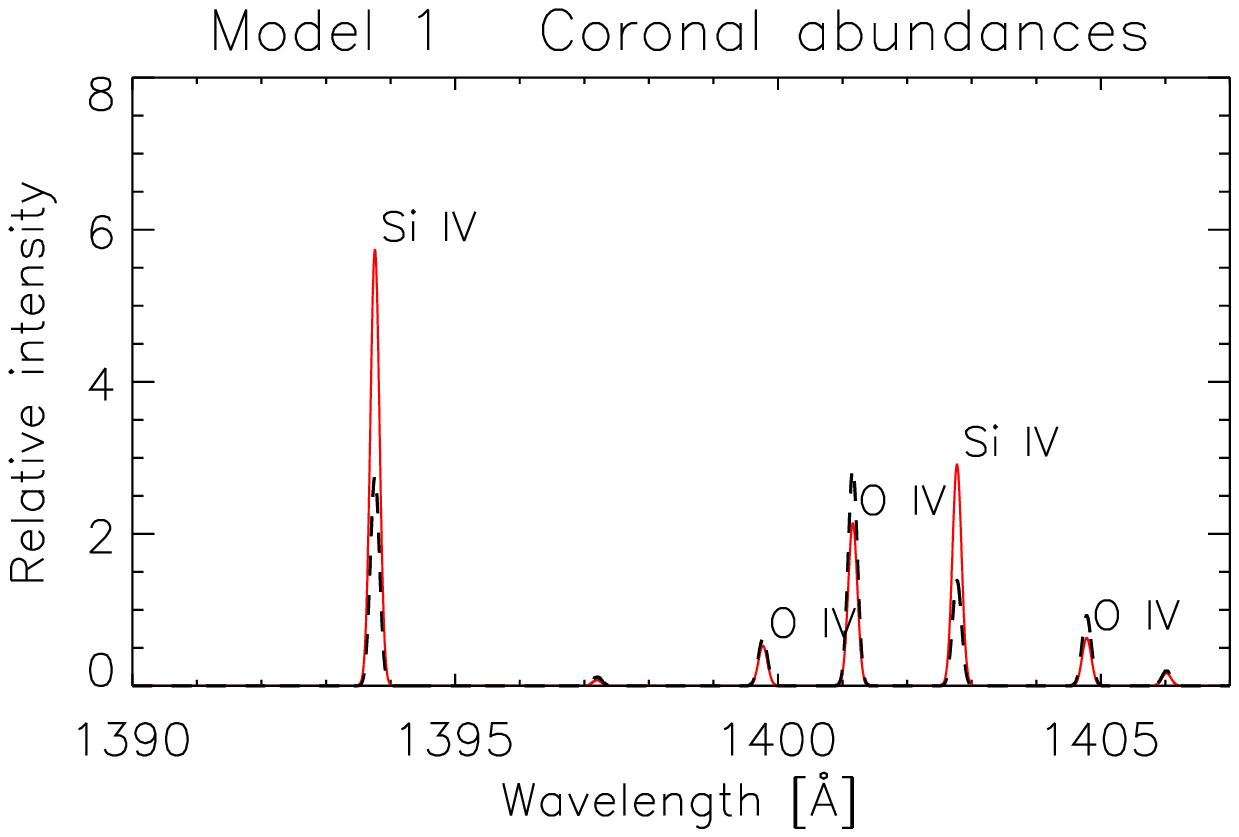}}
  \resizebox{\hsize}{!}{\includegraphics{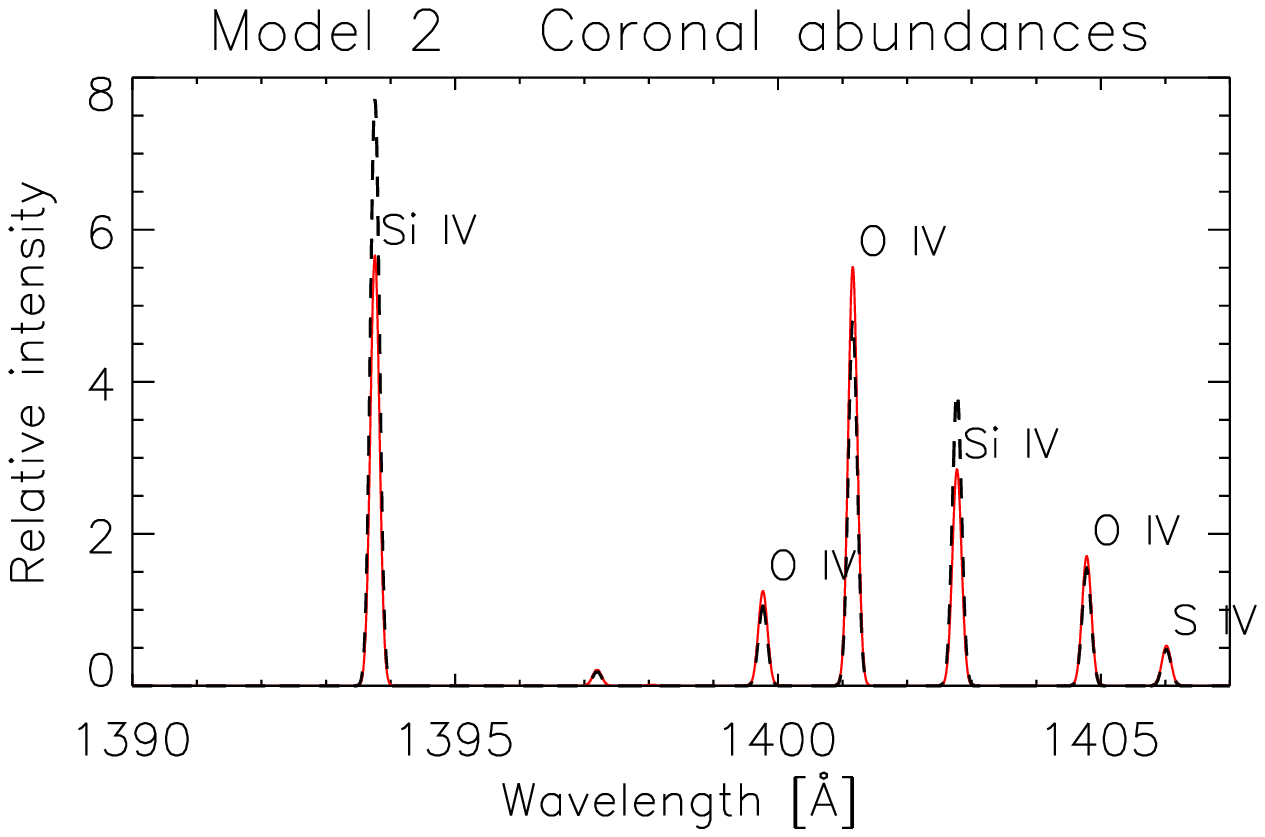}}
  \caption{Calculated synthetic spectrum observable in IRIS with the coronal abundances ({\it full red lines}) for Model 1 ({\it top}) and Model 2 ({\it bottom}) in a comparison with the Maxwellian spectrum for the same temperature and density structure of the loop ({\it dashed black lines}). }
  \label{Fig:spectrum_cor}
\end{figure}

%
%__________________________________
\section{Model transition region}
\label{Sect:2}

Suprathermal transition region electron velocity distribution functions were calculated in Paper I by solving the Boltzmann-Vlasov equation including Coulomb collisions. This kinetic model is described in detail in   Paper I. It is based on the coronal loop model of \citet{Vocks08}.  The loop is 210\,Mm long and its geometry is given by a potential extrapolation of a photospheric magnetogram observed on 2003 October 28. The loop has an apex temperature of 1.4\,MK and a footpoint electron density of 2\,$\times$\,10$^9$ cm$^{-3}$. The new   transition region simulation box is located below that of \citet{Vocks08}, and the electron distribution at the loop footpoint in \citet{Vocks08} is used as an upper boundary condition for the transition region model.

This model requires a background fluid model, that is,~densities and temperatures, for both electrons and ions in the transition region. The temperature profile is based on the classic Spitzer $T^{5/2}$ law of thermal conductivity in a plasma, starting from the temperature at the upper boundary. The
  density profile is then calculated by a hydrostatic model.

In order to understand the effect of transition region thickness, a second model with an artificially thick transition region, based on a $T^{3/2}$ law
of thermal conductivity, has been prepared.  The upper boundary  conditions for density, temperature, and electron distribution, were not
  changed. This modification does not consider possible implications on the  plasma state inside the loop, as they are discussed by
~\citet{Peter12}, for example.

From now on, the transition-region models based on the $T^{5/2}$ and $T^{3/2}$ thermal conductivity are simply referred to as Models 1 and 2, respectively. The resulting profiles of $T$ and $N_\mathrm{e}$ are shown in Fig. \ref{Fig:model}.  The increased  thickness of the transition region in Model 2 can readily be seen, as the  temperature reaches values of $1.0\times10^{5}$ K or  $2.0\times10^{5}$ K, for example, at a larger height for Model 2 than for Model 1.

The resulting electron distributions as a function of energy are shown in Fig. \ref{Fig:distr} for both models. It can be seen that there are strong suprathermal tails present in the transition region, although the total numbers of the particles in the high-energy tails are small in comparison with the bulk of distribution in both Models. These high-energy tails arise as a result of collisionless particles streaming down the transition region from the hotter corona, where non-Maxwellian particle populations arose due to resonant interaction with the whistler waves, as discussed by \citet{Vocks08}.

The high-energy tails in Model 1 are typically approximately one order of magnitude higher than in the Model 2 for electron energies of several 10 eV up to a few 100 eV {\em (Fig. \ref{Fig:distr} and Tab. \ref{Tab1}).} At higher electron energies, the difference becomes smaller, as electrons with such energies can traverse the transition region essentially collision-free. The relative number of the electrons in the high-energy tail reaches its maximum $4\%$ at log($T$) $\approx4.6$ for Model 1. It is important that these electrons carry approximately $25\%$ of the total energy of the electron distribution (Tab. \ref{Tab1}, {\it left}) and this energy source can have a significant effect on ionization and excitation. On the other hand, the relative number of the electrons in Model 2 is typically bellow $1\%$  (Tab. \ref{Tab1}, {\it right}). The high energy tail contains a few \% of the total energy only and its effect is smaller than in Model 1. 
\begin{table}
\caption{ The  number of particles and their energy in the high-energy tail relative to the total particle number and the energy of the electron distribution for Model 1 ({\it left}) and Model 2 ({\it right}) as a function of temperature in the transition region. }
\begin{tabular}{| c c c | c c c |}
\hline
\multicolumn{3}{|c|}{Model 1} & \multicolumn{3}{c|}{Model 2}\\
Log(T) & particles &  energy & Log(T) & particles &  energy \\
 $ $[K] &  [\%] &  [\%]  &  $ $[K] &  [\%] &  [\%]  \\
\hline
4.05  & 0.9  & 5.5  & 4.05   &  0.9   &   1.6 \\
4.16  & 2.2  &  25 &   &   &   \\ 
4.45  & 3.3 & 26  &  4.47  & 1.6   &   2.3 \\
4.67 & 3.8  &  22  & 4.71 &  0.7   &   2.3 \\
4.82 & 3.4 &   20  & 4.84 & 0.4   &  1.9 \\
5.01 &  2.5  &   14   & 4.98  & 0.3   &   1.7 \\
 5.23 &  1.7  &  8.2  & 5.26  & 0.3   &   1.6 \\
 5.37 &  1.2 &   5.6  & 5.36  &  0.3  &   1.4 \\
 5.52 &  0.8  & 3.1 &  5.57  &  0.2   &  1.1  \\
 5.74 &  0.4 &  0.6   & 5.75  &  0.2   &  0.6\\
\hline
\end{tabular}
   \label{Tab1}
\end{table}
%
%
%__________________________________
\section{Synthetic spectra}
\label{Sect:3}

Using the models described in Sect. \ref{Sect:2}, and especially the resulting distribution functions, we have calculated the ionization and excitation equilibrium, as well as the synthetic spectra for Si, O, and S at the spectral range 1390--1407 \AA. This spectral range was chosen to model the transition region line emission observed by {IRIS} \citep[\textit{IRIS},][]{DePontieu14}.

The line intensity $I$ formed between levels $i$ and $j$ in the optically thin plasma is the integral of the emissivity, $\varepsilon_{ij}$, along the line of sight
\begin{equation}
I=\frac{1}{4\pi}  \int_{l} \varepsilon_{ij}dl=\frac{hc}{4\pi\lambda_{ij}} A_{ij} A_{\mathrm{X}} \int_{l} \frac{N^{+k}_{\mathrm{X},i}}{
N^{+k}_{\mathrm{X}}} \frac{N^{+k}_{X}}{N_{\mathrm{X}}}\frac{N_{\mathrm{H}}}{N_{\mathrm{e}}} N_{\mathrm{e}} dl,
\end{equation}
where $h$ is Planck constant, $\lambda_{ij}$ is the line wavelength corresponding to the transition from atomic level $i$ to level $j$, $A_{ij}$ is the corresponding Einstein coefficient for spontaneous emission, $N^{+k}_{\mathrm{X},i}$ is the number of ions with the excited level $i$, $N^{+k}_{\mathrm{X}}/N_{\mathrm{X}}$ is the relative abundance of $+k$-times ionized ions to the total number of ions for element $\mathrm{X}$, $A_{\mathrm{X}}$ is the relative
abundance of the element $\mathrm{X}$ to hydrogen, $N_{\mathrm{H}}$ is the total number of hydrogen ions, and $N_{\mathrm{e}}$ is the electron density.  In equilibrium, the ratio $N^{+k}_{\mathrm{X},i} / N^{+k}_{\mathrm{X}}$ is given by the excitation equilibrium and $N^{+k}_{\mathrm{X}}/N_{\mathrm{X}}$ is given by the ionization one. 

\subsection{Ionization equilibrium}
\label{Sect:3.1}

In the high-temperature and low-density plasma, the electron direct ionization and autoionization together with radiative and dielectronic recombination are the dominant ionization and recombination processes. As such, they are important for the ionization equilibrium calculation. The rate of an elementary process for any energy distribution function, $f({\cal E})$, can be written
\begin{equation}
        R=\langle\sigma v\rangle=\int_{0}^{\infty} \sigma f({\cal E})~\left(\frac{2{\cal E}}{m}\right)^{1/2} \mathrm{d} {\cal E} 
        \label{Eq:ioniz_rate}
,\end{equation}
where $\sigma$ is the cross section, $v$ is the electron velocity, and $\cal E$ is the corresponding electron energy. The cross sections for the direct ionization and autoionization calculation were taken from the CHIANTI database, version 8.0 \citep{DelZanna15}. They  correspond to atomic data of \citet{Dere07}. \citet{Dere09} used these data to calculate the ionization equilibria for the Maxwellian distribution. The ionization rates given by Eq.~(\ref{Eq:ioniz_rate}) for our model distributions have been calculated numerically. As a result of the presence of the high-energy tails of the distribution, the ionization rates show a strong increase at low temperatures (Fig. \ref{Fig:rates}). The effect is more pronounced for Model 1 where more high-energy particles are present (see Fig. \ref{Fig:distr}).

For the radiative recombination, we used the method of \citet{Dzifcakova92}, \citet{Wannawichian03}, and \citet{Dzifcakova13}. Although this method was developed for the $\kappa$-distributions, it can be used for any distribution, including numerical ones. The cross-section for the radiative recombination, $\sigma_\mathrm{RR}$, is assumed to have the form \citep[e.g.,][]{Osterbrock74}
\begin{equation}
\sigma_{\mathrm{RR}}=C_{\mathrm{RR}}/{\cal E}^{\eta+0.5}
,\end{equation}
where $C_{\mathrm{RR}}$ is a constant and $\eta+0.5$ is a power-law index. Both parameters can be found in \citet{Aldrovandi73} \citet{Landini90}, \citet{Shull82}, \citet{Mazzotta98}, or \citet{Badnell06}, for example. It should be noted that the recombination cross-section decreases with the incident energy and low-energy electrons are the main contributors to the radiative recombination rates. Therefore, the radiative recombination rates for our model distributions are nearly the same as the Maxwellian rates at temperatures corresponding to the distribution bulks. The small numbers of high-energy particles only have a minor effect on these rates \citep[see also][]{Roussel-Dupre80b}.

Dielectronic recombination rates can be approximated by the following expression valid for any distribution function \citep{Dzifcakova92,Dzifcakova13}  
\begin{equation}
R_{\mathrm{DR}}=\sum_m c_m \frac{\pi ^{1/2}}{2} \frac{f({ E}_m) }{{ E}_m^{1/2}}
,\end{equation}
where $c_m$ and ${ E}_m$ are the parameters given by the approximations of the Maxwellian dielectronic recombination rates. We have taken data from \citet{Abdel-Naby12}  \citet{Aldrovandi73}, \citet{Altun04}, \citet{Altun06}, \citet{Altun07}, \citet{Colgan03}, \citet{Colgan04}, \citet{Mitnik04}, \citet{Zatsarinny04}, \citet{Zatsarinny05a}, \citet{Zatsarinny05b}, and \citet{Zatsarinny06}. These data are also part of the CHIANTI database since version 6 \citep[see][]{Dere09}.

In equilibrium, direct ionization and autoionization are compensated by the radiative with dielectronic recombination, leading to the following relation
\begin{equation}
\frac{N^{+k}_{\mathrm{X}}}{N^{+k+1}_{\mathrm{X}}} =\frac{R_{\mathrm{DI}}+R_{\mathrm{AI}}}{R_{\mathrm{RR}}+R_{\mathrm{DR}}}
,\end{equation}
where $R_{\mathrm{DI}}$ is the rate of direct ionization and $R_{\mathrm{AI}}$ is the rate of autoionization.
The ionization equilibria for the model distributions were calculated in selected points along the loop. The abundace of \ion{Si}{II}--\ion{Si}{VI} and \ion{O}{I}--\ion{O}{VI} ions as a function of temperature along the loop are shown in the Figs. \ref{Fig:ioneq_si} and \ref{Fig:ioneq_o}. For comparison, relative abundances of the these ions for the Maxwellian distributions with the same temperatures are shown by dashed lines. It can be seen that the relative ion abundances are very sensitive to the presence of the high-energy tail. The ions in our models can exist in a wider temperature range in comparison with the Maxwellian case. In particular, the ions can also exist at very low temperatures. Furthermore, the maxima of the relative ion abundances are shifted to the lower temperatures. This shift is much stronger for the Model 1 where the number of high energy particles is higher. The behavior of the ionization peaks for our model distribution is, in general terms, similar to the effect of the $\kappa$-distributions on the ionization equilibrium in the transition region \citep{Dzifcakova13}, although the details obtained here are different. The relative ion abundances calculated here show that even relatively weak non-Maxwellian tails in the transition region will have a significant impact on the ionization equilibrium, as also qualitatively discussed already by \citet{Roussel-Dupre80b}.
\subsection{Excitation equilibrium}
\label{Sect:3.2}

In the solar transition region, the dominant processes are the electron collisional excitation and deexcitation together with the spontaneous radiative decay transitions. Photoexcitation can be negleted at the electron densities in our models \citep{Dudik14a}. Generally, the electron excitation rates are calculated from the non-dimensionalized collision strengths $\Omega_{ij}$ instead of the cross-sections $\sigma_{ij}$ themselves. These two quantities are related by
\begin{equation}
        \sigma_{ij} = \frac{\Omega_{ij}}{ \omega_{i} {\cal E}_{i}}\pi a_{0}^{2}\,,
        \label{Eq:Omega}
\end{equation}
where $E_{ij}$ is the excitation energy, $\omega_{i}$ is the statistical weight of the level $i$, ${\cal E}_{i}$ is the incident electron energy, $I_\mathrm{H}$ is the hydrogen ionization energy, and $a_{0}$ is the Bohr radius.

The CHIANTI database contains Maxwellian-averaged collision strengths or their spline approximations for the majority of the astronomically interesting ions of elements from H to Zn \citep{Dere97,DelZanna15}. However, CHIANTI software enables calculation of spectra for the Maxwellian plasma only. Consequently, \citet{Dzifcakova15} developed a method to calculate $\Omega$ for the approximate excitation and de-excitation rates for non-Maxwellian distributions. They tested and applied this method for the $\kappa$-distributions and showed that the approximate method yields collisional excitation rates within 5--10\% compared to the direct numerical calculations. These approximations of $\Omega$ are contained in the KAPPA package \citep{Dzifcakova15}. For our model distributions (Sect. \ref{Sect:2}), we used these approximations to calculate excitation equilibria and synthetic spectra of \ion{Si}{IV}, \ion{O}{IV,} and \ion{S}{IV} in the \textit{IRIS} FUV wavelength window. These calculations are based on the atomic data of \citet{Liang09b}, \citet{Liang09a}, \citet{Feuchtgruber97} ,\citet{Liang12}, \cite{Correge04}, \cite{Foster97}, \citet{Tayal00}, \cite{Tayal99}, \citet{Hibbert02}, \citet{Johnson86}, and \citet{Bely70}. 
\subsection{Line intensities}
\label{Sect:3.3}

Having obtained the ionization and excitation equilibria (Sects. \ref{Sect:3.1} and \ref{Sect:3.2}, respectively), we next calculated the emissivities of the \ion{Si}{IV} and \ion{O}{IV} lines for both models. Since the amount of high-energy particles present in the model transition region is small (Sect. \ref{Sect:2} and Table \ref{Tab1}) despite their effects on ionization equilibrium, the synthetic spectra obtained here are compared with the Maxwellian ones. We remind the reader that the Maxwellian spectra typically show higher \ion{O}{IV} 1401.2\,\AA~intensities compared to \ion{Si}{IV} 1402.8\,\AA~ones. This is contrary to the observed spectra, where \ion{Si}{IV} is enhanced compared to \ion{O}{IV}, by a factor of five or more. Thus, rather than reproducing the observed spectra in detail, our aim here is only to investigate the effect (if any) of the non-Maxwellian particles within Models 1 and 2 on the emergent transition region spectra.

The calculated emissivities of \ion{Si}{IV} and \ion{O}{IV} lines for both models are shown in Figs. \ref{Fig:emiss_si} and \ref{Fig:emiss_o}. It can be seen that even a small number of high-energy particles results in the enhancement of emissivity by several orders at temperatures close to $10^{4}$\,K for Model 2. Both the \ion{Si}{IV} and \ion{O}{IV} contribution functions have a pronounced wing towards temperatures as low as 3$\times$10$^{4}$\,K. Increase of the high-energy particles in Model 1 results in formation of \ion{Si}{IV} and \ion{O}{IV} emissivity maxima at temperatures of approximately  $1.5\times10^{4}$ K, which is approximately one order of magnitude lower than the temperature of emissivity maxima in the Maxwellian case, and also significantly lower than in Model 2.

The synthetic spectra in the \textit{IRIS} FUV channel are shown in red in Figs. \ref{Fig:spectrum_pho} and \ref{Fig:spectrum_cor}, where the corresponding Maxwellian spectra are also shown by black dashed lines. The spectra are integrated along the model atmosphere shown in Fig. \ref{Fig:model}. This is since the transition region here corresponds to a footpoint of a coronal loop (Paper I) rather than a purely transition-region structure such as a small loop. While the synthetic spectra are obtained by using the full modeled distribution functions (Fig. \ref{Fig:distr}), the corresponding Maxwellian spectra are obtained using only the Maxwellian cores of the distribution at each location. Finally, the spectra shown in Fig. \ref{Fig:spectrum_pho} were obtained using the photospheric abundances \citep{Asplund09}, while those in Fig. \ref{Fig:spectrum_cor} correspond to the coronal abundances \citep{Feldman92}.

The synthetic spectra reflect the behavior of the emissivities shown in Figs. \ref{Fig:emiss_si} and \ref{Fig:emiss_o}. For Model 2, the resulting spectrum is similar to the Maxwellian spectrum (Fig. \ref{Fig:spectrum_pho} and \ref{Fig:spectrum_cor}, bottom); however, the \ion{O}{IV} intensities are weakly enhanced compared to the Maxwellian, while the \ion{Si}{IV} are depressed by around 20--30\%. The \ion{S}{IV} is almost unaffected. Such changes in the \ion{O}{IV} and \ion{Si}{IV} intensities depart further from the typically observed spectra. Thus, Model 2 does not help in reducing the discrepancy between the synthetic and observed \ion{Si}{IV} and \ion{O}{IV} intensities.

On the other hand, for Model 1, the \ion{Si}{IV} 1402.8\,\AA~line is increased by more than a factor of 2 compared to the Maxwellian case, while the \ion{O}{IV} intensities are depleted by several tens of per cent (Table \ref{Tab2} and top panels in Figs. \ref{Fig:spectrum_pho}--\ref{Fig:spectrum_cor}). 
This behavior is qualitatively similar to the effect of the $\kappa$-distribution on the \textit{IRIS} spectra \citep{Dudik14a}. In our case however, the decrease of \ion{O}{IV} relative to \ion{Si}{IV} is not so pronounced, and the \ion{Si}{IV} to \ion{O}{IV} intensity ratios change only by approximately a factor of 2. This comes from the much weaker high-energy tail in our model distribution (Fig. \ref{Fig:distr}) compared to a $\kappa$-distribution. Nevertheless, the spectra obtained for Model 1 help to reduce the discrepancy between the synthetic Maxwellian and the actually observed ones. This shows that even a relatively small number of energetic particles, generated with a simple model of the transition region, can still have a significant effect on the emergent spectra.
\begin{table}
\caption{Relative intensities of  \ion{O}{iv} 1401.16 \AA\,and 1399.77 \AA\,line to \ion{Si}{iv} 1393.76 \AA\,line for Model 1({\it left}) and Model 2 ({\it right}) assuming photospheric and coronal abundances. }
\begin{tabular}{| c | c c | c  c |}
\hline
\multicolumn{1}{|c}{ } & \multicolumn{2}{|c|}{Model 1} & \multicolumn{2}{c|}{Model 2}\\
 \ion{O}{iv} & 1401.16 \AA &  1399.77 \AA  &  1401.16 \AA & 1399.77 \AA   \\
\hline
\multicolumn{1}{|c}{ } & \multicolumn{4}{|c|}{ Photospheric abundances } \\ 
Model     &  0.94  &  0.23  &  2.40   &  0.55    \\
Maxwell  &  2.60  &  0.56  &  1.56   &    0.34   \\ 
 \hline
 \multicolumn{1}{|c}{ } & \multicolumn{4}{|c|}{ Coronal abundances } \\ 
Model     &  0.37  &  0.09  & 0.98   &  0.22    \\
Maxwell  &  1.04  &  0.22 &  0.63   &   0.14   \\ 
 \hline 
\end{tabular}
   \label{Tab2}
\end{table}

%__________________________________
\section{Summary and Discussion}
\label{Sect:4}

We calculated the synthetic transition-region spectra emerging from the transition region model of Paper I. This model considers the propagation of particles in the transition region below a coronal loop, where high-energy tails are created by whistler-wave turbulence \citep{Vocks08}. The high-energy tails are relatively weak, containing less than 4\% of the particles. These high-energy tails enhance the ionization rates, which show pronounced low-temperature shoulders. Consequently, ions such as \ion{Si}{III}--\ion{Si}{V} and \ion{O}{III}--\ion{O}{V} exist also at very low temperatures almost down to 10$^4$\,K. This behavior is in turn translated to the behavior of the contribution functions of the \ion{Si}{IV} and \ion{O}{IV} lines observed by \textit{IRIS}. The resulting synthetic spectra for the Model 1 show increase of the \ion{Si}{IV} intensities by a factor of more than 2, while the \ion{O}{IV} lines are weakly decreased when compared to the corresponding Maxwellian spectra. For Model 2, the \ion{O}{IV} lines are weakly enhanced as a result of their contribution functions being more flat at transition region temperatures than the \ion{Si}{IV} ones.

We note that in both cases, the synthetic spectra obtained here still contain \ion{O}{IV} 1401.2\,\AA~intensities comparable to the \ion{Si}{IV} 1402.8\,\AA~ones, while in observations, the \ion{Si}{IV} intensities are much larger. Transient ionization has been shown to strongly affect the \ion{Si}{IV} and \ion{O}{IV} intensities. The spectra calculated from models of solar atmosphere, including transient ionization \citep[e.g.,][]{Doyle13,Olluri13a,Olluri13b,Martinez16}, show enhancements of \ion{Si}{IV} relative to \ion{O}{IV} similar to observations. Although the importance of the transient ionization is beyond reasonable doubt, our results show that the propagation of high-energy particles through the transition region also needs to be considered as it can also lead to enhancements of the \ion{Si}{IV} lines and weak depletions of the \ion{O}{IV} ones when compared to the equilibrium Maxwellian spectrum. We note that the high-energy tails considered here occur naturally in the transition region. Thus, these high-energy tails exist even outside reconnection events or strong electric fields, which would likely increase the high-energy tails and further modify the emergent spectrum.

\begin{acknowledgements}
This work has been supported by Grant Nos. 16-18495S and 17-16447S of the Grant Agency of the Czech Republic. We used the CHIANTI software and database. CHIANTI is a collaborative project involving the NRL (USA), RAL (UK), MSSL (UK), the Universities of Florence (Italy) and Cambridge (UK), and George Mason University (USA). The authors benefited greatly from participation in the International Team 276 funded by the International Space Science Institute (ISSI) in Bern, Switzerland.
\end{acknowledgements}

\bibliographystyle{aa}
\bibliography{model_3}

\end{document}